\documentclass{JHEP3}

\preprint{LU-TP 07-07\\
  hep-ph/0702241}

\usepackage{cite}
\usepackage{epsfig}
\usepackage{color}

\usepackage{graphics}
\usepackage{axodraw}
\usepackage{inputenc}
\usepackage{xspace}
\inputencoding{latin1}
 \renewcommand\email[1]{{\scriptsize\tt\href{mailto:#1}{#1}}}

\newcommand{\pythia}{P\scalebox{0.8}{YTHIA}\xspace}

\def\mrm#1{\mathrm{#1}}
\def\sub#1{\ensuremath{_{\mrm{#1}}}}
\def\sup#1{\ensuremath{^{\mrm{#1}}}}

\newcommand{\done}[1]{}

\newcommand{\eqref}[1]{eq.~(\ref{#1})\xspace}

\skip\footins = 1\bigskipamount plus 2pt minus 4pt                              

\title{\boldmath String Effects on Fermi--Dirac Correlation Measurements}

\author{Rosa Mar\'ia Dur\'an Delgado, Gösta Gustafson and Leif Lönnblad\\
  Dept.~of Theoretical Physics,
  S\"olvegatan 14A, S-223 62  Lund, Sweden\\
  E-mail: \email{rositaduran@yahoo.es}, \email{Gosta.Gustafson@thep.lu.se}
    and \email{Leif.Lonnblad@thep.lu.se}}
  
  \abstract{We investigate some recent measurements of Fermi--Dirac
    correlations by the LEP collaborations indicating surprisingly
    small source radii for the production of baryons in
    $e^+e^-$-annihilation at the $Z^0$ peak. In the hadronization
    models there are besides the Fermi--Dirac correlation effect also a
    strong dynamical (anti-)correlation. We demonstrate that the
    extraction of the pure FD effect is highly dependent on a
    realistic Monte Carlo event generator, both for separation of
    those dynamical correlations which are not related to Fermi--Dirac
    statistics, and for corrections of the data and background
    subtractions. Although the model can be tuned to well reproduce
    single particle distributions, there are large model-uncertainties
    when it comes to correlations between identical baryons. We
    therefore, unfortunately, have to conclude that it is at present
    not possible to make any firm conclusion about the source radii
    relevant for baryon production at LEP.}

\begin{document}

\section{Introduction}
\label{sec:introduction}

Hanbury-Brown and Twiss used Bose--Einstein (BE) correlations between
photons to measure the size of distant stars
\cite{HanburyBrown:1956pf}. A pair of bosons produced incoherently
from an extended source will have an enhanced probability, $P_{12}$,
to be found close in momentum space when detected simultaneously, as
compared to if they are detected separately ($P_1$, $P_2$). If the
production region has a Gaussian shape with some radius, $R$, it is
fairly easy to show that the enhancement is given by
\begin{equation}
  \label{eq:BEbasic}
  C\sub{BE}\equiv\frac{P_{12}}{P_1P_2}=1+e^{-Q^2R^2},
\end{equation}
where $Q^2=-(k_1-k_2)^2$ is the negative of the square of their
four-momentum difference. 

It was also early proposed to use a similar analysis to gain
information on the geometry of the production region for pions in
high-energy collisions \cite{Goldhaber:1960sf}. The assumption of
completely incoherent production in a Gaussian region is obvious when
considering photons from a star, and also very reasonable for the
production of pions in central heavy-ion collisions. For hadronic
collisions or $e^+e^-$ annihilation it may, however, seem much less
natural. The assumption of incoherent production implies that the
source is undisturbed by the emission, and thus not affected by the
enhanced radiation. In $e^+e^-$ annihilation the source disappears in
the hadronization process. Energy--momentum conservation is an
important constraint, and the source is not even approximately
constant. The distribution of pions is also far from isotropic,
usually concentrated in narrow jets, and further complicated by the
fact that the pions often come from decays of long-lived heavier
resonances.  In spite of all these problems, introducing a so-called
caoticity parameter in eq.~(\ref{eq:BEbasic}), and still assuming a
Gaussian production region, all $e^+e^-$ experiments arrive at a
remarkably consistent value for the size of the production region:
$R\approx0.5-1$~fm.

With a confining force, or string tension, of the order 1~GeV/fm this
might have been regarded as a small production region for collisions
at e.g.\ LEP energies where the $q\bar{q}$ pair is separated by about
90~fm before they are stopped. However, in successful models based on
strings or cluster chains, hadrons which are close in momentum space
originate from regions which are also close in coordinate space.
Although the origin for the correlation is not fully understood, a
production region for pions or kaons of the order of 1~fm is therefore
quite reasonable.

An attempt to explain the observed correlation as an effect of quantum 
interference between different contributions to the production
amplitude in the string hadronization process is presented in 
refs.~\cite{Andersson:1985qn, Andersson:1997hs}. Although this 
approach gives a qualitatively correct
result, quantitative predictions at LEP energies have been hampered
by technical difficulties. Within this approach it has been argued that the 
correlation between string pieces separated by a gluon should be strongly 
reduced. When the center-of-mass energy is increased the number of gluons is 
also large, and the mass of a straight string piece between two 
gluons is kept relatively small. The result is therefore sensitive 
to the hadronization of small string systems.
The iterative solution to the Lund string hadronization model
\cite{Andersson:1983jt} is only exact for high mass systems, and 
although the corrections due to finite energy normally can be neglected for 
\emph{inclusive distributions}, they do have a large impact on 
\emph{correlations} \cite{Sandipanprivate}.

\done{BE correlations in stars and multi-particle production,
  measuring the size of the production region}

A natural parallel to BE correlations is to look at the corresponding
correlation between identical fermions. Here one would expect a
depletion of fermion pairs close in momentum space, and using the same
assumptions for the production region as above we arrive at
\begin{equation}
  \label{eq:FDbasic}
  C\sub{FD}=1-\lambda e^{-Q^2R^2},
\end{equation}
where we now have explicitly included the ``caoticity parameter''
$\lambda$. Recently three of the LEP experiments have published
results on such Fermi--Dirac (FD) correlations for $pp$,
$\bar{p}\bar{p}$, $\Lambda\Lambda$ and $\bar{\Lambda}\bar{\Lambda}$,
again finding consistent results giving $R\sim0.15$~fm. This result is a
bit disturbing, not only because the size of the production is much
smaller than the one obtained for mesons, but also because the size is
smaller than the baryons themselves. It is therefore important to
thoroughly investigate possible theoretical and/or experimental problems 
in the analyses.

\done{FD correlations could be an important cross-check}

\done{Measurments indicate a production region for baryons, smaller than
  that of mesons, and smaller than the size of the baryons
  themselves.}

An essential problem in extracting the correlation is the estimate
of the reference distribution $P_1P_2$ in eq.~(\ref{eq:BEbasic}).
The distributions $P_1$ and $P_2$ must here correspond to exactly
identical events, i.e.\ events with the same emission of gluon
radiation (and the same orientation). If there are other correlations
beside the BE or FD effect, the distribution $P_1P_2$ should be 
replaced by a reference distribution 
corresponding to the two-particle distribution in a hypothetical world
without BE or FD correlations. If $N(Q)$ and $N_{\mathrm{ref}}(Q)$
represent the number of pairs in the real and the reference sample
respectively, the correlation is determined by the ratio
\begin{equation}
C(Q)=\frac{N(Q)}{N_{\mathrm{ref}}(Q)}.
\label{eq:CQ}
\end{equation}

Such a reference sample is obviously not directly observable in an
experiment. Different methods have been applied for the construction
of reference samples, but they all suffer from serious limitations.
One possibility is to use a phenomenological hadronization model
without BE or FD correlations, implemented in a Monte Carlo event
generator. Besides a strong model dependence this method is also
sensitive to any imperfection in the event-generator implementation of
the underlying hadronization model.

To reduce the model sensitivity it is therefore preferable to
construct reference samples directly from data. Two different methods
have been used for this purpose: To use \emph{opposite sign pairs} and
to use pairs from \emph{mixed events}. When studying correlations in
$\pi^+\pi^+$ or $\pi^-\pi^-$ pairs it may seem reasonable to use a
reference sample of $\pi^+\pi^-$ pairs, which are free from the BE
effect. The $\pi^+\pi^-$ pairs have, however, strong correlations due
to resonance contributions, and when using the ratio
$N_{\pi^+\pi^+}(Q^2)/N_{\pi^+\pi^-}(Q^2)$ to determine the BE effect,
it is therefore necessary to cut out the resonance regions in the fit,
or else to estimate their contributions to the distribution. The
method to use a reference sample with pairs of particles from
different events (a mixed reference sample) has the problem that the
hadronization is dependent on the gluon radiation, which differs from
event to event. This bias can be reduced (but not eliminated) by a cut
in thrust, which limits the radiation, and by orienting the events to
align the thrust axes.

In this paper we want to discuss the special problems encountered when
trying to extract the effect of FD correlations in pairs of
(anti-)protons and $\Lambda$s.  Naturally there are great experimental
difficulties in a determination of $\bar{p}\bar{p}$ or
$\Lambda\Lambda$ correlations, which are associated with acceptance
limitations, an admixture of pions in the (anti-)proton sample and
limited statistics in the $\Lambda\Lambda$ sample.  We will here show
that the experimental analyses of momentum correlations also depend
very strongly on a realistic hadronization model, which can introduce
large errors due to uncertainties both in the model used and in its
implementation in \pythia\cite{Sjostrand:2003wg,Sjostrand:2006za}.  We
therefore conclude that it presently is not possible to confirm the
very small production regions presented in the literature.  It should
be mentioned that there is also a model-independent method to study
the FD effect, which is based on the spin correlation in
$\Lambda\Lambda$ pairs. This method is, however, limited by low
statistics, which also here prevents a definite conclusion.

The layout of the paper is as follows: In section
\ref{sec:meas-fermi-dirac} we describe how the experiments extract FD
effects in correlations between identical baryons. Then in section
\ref{sec:bary-prod-lund} we present the basics of the Lund string
hadronization model, where we in particular discuss baryon production
and baryon--baryon correlation. In sec.~\ref{sec:uncertainties} we
discuss in more detail the uncertainties in the Lund model and the
approximations in its implementation in the \pythia event generator,
which have an impact on the baryon correlations. Finally our main
results are summarized in sec.~\ref{sec:conclusions}.

\section{Measurements of Fermi--Dirac correlations}
\label{sec:meas-fermi-dirac}

As mentioned in the introduction there are two different method, 
which have been used to determine
the FD correlations between identical baryons. One is based on 
momentum correlations similar to the analyses of BE correlations 
between meson pairs discussed above, while the other uses
spin correlations in $\Lambda\Lambda$ pairs. While the first is 
model dependent the second is hampered by low statistics. We
will in this section discuss the results of both methods.

\subsection{Momentum correlations}
\label{sec:mom-corr} 

Experimental measurements (and theoretical expectations) show strong
correlations between a proton and an anti-proton or between a $\Lambda$
and an $\bar\Lambda$.
In the recent measurements of FD correlations between protons and
$\Lambda$s, the reference samples have therefore typically been constructed using
pairs of particles coming from different events, rather than opposite charge
particles.
To reduce the bias due to gluon radiation the events have frequently
been oriented so that the thrust axes end event planes coincide, and
sometimes also a lower cut on the thrust value has been applied. 

Associated with the baryon--anti-baryon correlations there are in
current hadronization models also very strong correlations between
identical baryons, besides those caused by gluon emission.  These
correlations must be separated before the true FD effect can be
extracted. A way to reduce the bias due to both the gluon radiation
when using a mixed reference sample, and the dynamical correlations
which are not due to FD statistics, is to compare with expectations
from a Monte Carlo (MC) simulation program. This is usually done by
studying the "double ratio"
\begin{equation}
  \label{eq:BEdouble}
  C(Q)=\left(\frac{N^{\mathrm{exp}}(Q)}
  {N_{\mathrm{ref}}^{\mathrm{exp}}(Q)}\right)\bigg/
  \left(\frac{N\sup{MC}(Q)}{N_{\mathrm{ref}}\sup{MC}(Q)}\right).
\end{equation}
Here $N\sup{MC}$ and $N_{\mathrm{ref}}\sup{MC}$ represent the MC
generated real pairs and pairs from a generated reference sample of mixed
events. With a realistic MC this method should isolate the true FD effect.

We want to emphasize that this method necessarily has the problem,
that the result is very sensitive to the description of the
correlations in the MC, which should be free of FD effects. These
correlations are much more uncertain than the inclusive distributions,
which have been accurately tuned to experimental data (see e.g.\
\cite{Hamacher:1995df}). Note also that they cannot be constrained by
data, as it is not possible to switch off FD effects in the
experiment.  It is exactly the difference between the data and the
FD-free MC, which is interpreted as the FD effect. If the model and/or
its MC implementation is imperfect the result will be wrong.  We here
also note that if the MC is tuned to correctly describe the single
proton spectrum, then it automatically also reproduces the pairs in
the mixed events. This implies that the double ratio in
eq.~(\ref{eq:BEdouble}) is actually very close to the single ratio,
with the MC result as the reference sample in eq.~(\ref{eq:CQ}). As
far as we know, there has been no measurements of correlations between
non-identical baryons, e.g.\ $p\Lambda$ and $\bar{p}\bar{\Lambda}$.
Such a measurement would be a very effective tool for validating the
baryon--baryon correlations in the hadronization models in the absence
of FD effects.

The method with double ratios has been used in analyses by the three LEP 
experiments ALEPH, OPAL, and DELPHI. Their results are presented in 
table \ref{table}, and we see that they all find similar results
with a production radius of the order of 0.15~fm.

\TABLE{
  \begin{tabular}{c c c c l}
    \hline
    & \textbf{$R \: (fm)$} & \textbf{$\lambda$} & \textbf{Experiment}\\
    \hline
    $\bar{p}\bar{p}$ & $0.14 \pm 0.06$ & $0.76 \pm 0.33$ & OPAL\footnotemark[1]
    & \cite{opal} \\
    & $0.11 \pm 0.01$ & $0.49 \pm 0.09$ & ALEPH  & \cite{Schael:2004qn} \\
    & $0.16 \pm 0.05$ & $0.67 \pm 0.25$ & DELPHI & \cite{delphi1,delphi2}\\
    \hline 
    $\Lambda\Lambda$ & $0.11 \pm 0.02$ & $0.59 \pm 0.10$ & ALEPH  &
    \cite{Barate:1999nv} \\
    & $0.17 \pm 0.14 $ & Spin Analysis &  ALEPH   & \cite{Barate:1999nv} \\
    & $0.19
    \begin{array}[c]{c}
      +0.37\\[-2mm]-0.07
    \end{array}
    $ & Spin Analysis &  OPAL   & \cite{Alexander:1996jq} \\
    & $0.11 
    \begin{array}[c]{c}
      +0.05\\[-2mm]-0.03
    \end{array}
    $ & Spin Analysis &  DELPHI\footnotemark[1]  & \cite{Lesiak:1998ij} \\
    \hline
  \end{tabular}
  \caption{\small{Experimental results for $\lambda$ and $R$ from FD
      correlations between baryon-pairs produced in $e^{+}e^{-}$
      annihilations at the LEP collider.}}
  \label{table} 
}

\footnotetext[1]{Note that these results has only been presented as a
  preliminary.}
\subsection{Spin--spin correlations}
\label{sec:spin-corr}

An alternative way to study the FD effect, which does not rely on
theoretical models and Monte Carlo simulations, is offered by the fact that
$\Lambda$ particles reveal their spin in the orientation of their decay 
products. A $\Lambda\Lambda$ pair with total spin 1 must
have an antisymmetric spacial wave function and is therefore expected to show 
a suppression for small relative momenta $Q$. $\Lambda\Lambda$ pairs with
total spin 0 has a symmetric spacial wave function, and should therefore
show an enhancement for small $Q$, similar to the correlation for bosons.
Therefore one expects pairs with small $Q$-values to be dominantly $S=0$.
This ought to be revealed in a preference for the protons from the 
decaying $\Lambda$s to be more back to back in the di-$\Lambda$ 
center-of-mass system for small $Q$.


Analyses at LEP \cite{Alexander:1996jq,Abbiendi:1998ux,Barate:1999nv}
do indicate such an effect. An example is the distribution in
$dN/dy^*$ as obtained by the ALEPH collaboration (figure 4 in
\cite{Barate:1999nv}). Here $y^*$ is the cosine of the angle between
two protons in the di-$\Lambda$ center of mass system. When fitted to
a straight line the resulting slope does increase for low $Q$-values,
thus favouring back-to-back correlations.  Results from three LEP
experiments are presented in table \ref{table}, and the fitted
production radii are consistent with those from analyses of momentum
correlations.  This type of fit to an expected form gives a very small
error, but looking at the result in \cite{Barate:1999nv}, this error
cannot represent the real uncertainty in the data, which can also be
well fitted by a horizontal line for all values of $Q$. Unfortunately
we have to conclude that the statistics is too limited for a reliable
determination of the range of the FD effect using this method.
(Although in principle model independent, also this method needs Monte
Carlo simulations to correct for losses due to acceptance and for
background contributions.)

\section{Baryon production in the Lund String Model}
\label{sec:bary-prod-lund}

The most successful model of hadron production is the Lund string
fragmentation model. In this model it is easy to show that two
identical baryons cannot be produced close in rapidity along a jet
and, hence, with small $Q$, since flavour-number conservation requires
that an anti-baryon is produced in between. This need not be the case
in other models. In e.g.\ the cluster hadronization model, two nearby
clusters may both decay isotropically into baryon--anti-baryon pairs
resulting in two identical baryons close in momentum space. It should
be noted that although the models can be tuned to reproduce inclusive
particle spectra with high accuracy, there are large uncertainties in
the description of particle correlations in general and baryon
correlation in particular.

As the Lund string model is used in the experimental analyses, we will 
here describe this model in some detail. The Lund model is
based on the assumption that the colour-electric field is
confined to a linear structure, analogous to a vortex line in
a superconductor. The model contains two basic components:
A model for the breakup of a straight force field and a model for a gluon
as an excitation on the string-like field. 

\subsection{Breakup by $q\bar{q}$ pair production.}

\FIGURE[t]{
  \includegraphics*[bb=160 300 460 500]{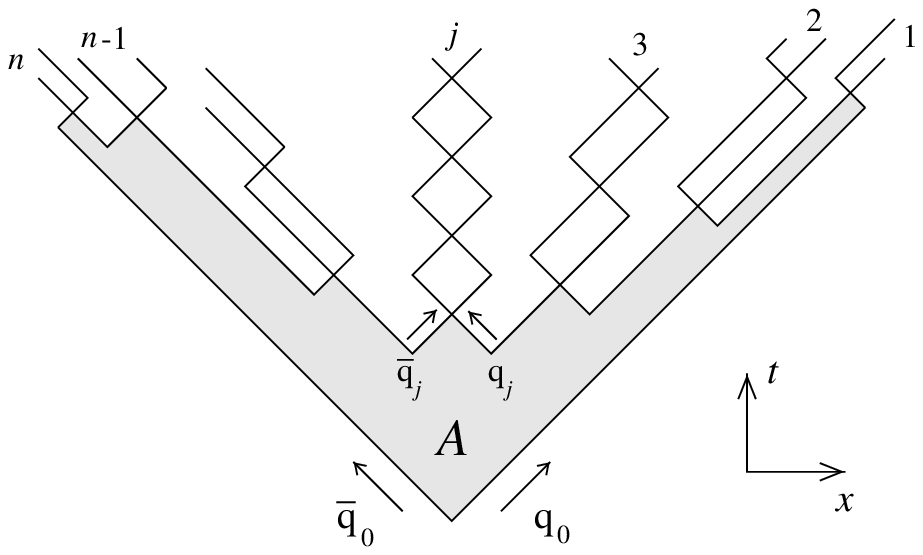}
  \caption{\label{fig:breakup} Schematic space--time picture of hadron
    production in the Lund string model.}}

The breakup of a high energy $q\bar{q}$ system is illustrated in 
fig.~\ref{fig:breakup}.
We study first a simplified one-dimensional world with only one meson mass 
and no baryon production. The probability for a final state
with $n$ mesons with mass $m$ and momenta $p_i$ is then given by the relation
\begin{equation}
  dP \propto \prod_i^n\Bigl[N d^2 p_i\delta(p_i^2 - m^2)\Bigr] 
  \delta\!\left(\sum p_i - P_{\mathrm{tot}}\right) \times \exp(-b A).
  \label{eq:breakup}
\end{equation}
Here $A$ is the space-time area indicated in fig.~\ref{fig:breakup}, while
$N$ and $b$ are two free parameters of the model.
The expression in eq.~(\ref{eq:breakup}) is a product of a phase space 
factor and the exponent of a "colour coherence area", $A$, which can be
interpreted as the imaginary part of an action.
The phase space is specified by the parameter $N$, where a large
$N$-value favours many particles and a small $N$ few particles.
The value of $b$ specifies the strength of the imaginary action,
and here a large value favours early breakups and correspondingly
few particles in the final mesonic state.

For a \emph{high energy system} the result in eq.~\ref{eq:breakup}
can be generated in an iterative way. The mesons can be "peeled
off" one by one from one end, where the $i$th meson takes the fraction
$z_i$ of the remaining (light-cone) momentum. The fractions $z_i$ 
are determined by the "splitting function"
\begin{equation}
f(z)=N\frac{(1-z)^a}{z} \exp(-\,\frac{b m^2}{z}).
\label{eq:f}
\end{equation}
Here the parameters $N$ and $b$ are the same as in
eq.~(\ref{eq:breakup}) (if measured in units such that the string
tension is 1) and $a$ is determined by the normalization constraint
$\int f(z) dz =1$.  (In practice, $a$ and $b$ are treated as the free
parameters and $N$ is determined from normalization.)

In the real world there are different quark species and different
meson masses. This is simulated by different weights $N_i$ in
eq.~(\ref{eq:f}), representing a suppression of strange quarks
and of the heavier vector mesons relative to pseudo-scalar mesons.
The parameter $b$ has to be a universal constant, but
$a$ can in principle vary depending on the quark flavour at the breakup, 
although most fits to data assume a single value. It is also necessary to 
include transverse momenta, where the meson mass, $m$, in 
eqs.~(\ref{eq:breakup}, \ref{eq:f}) is replaced by the transverse 
mass $m_\perp=\sqrt{m^2+p_\perp^2}$.

We emphasized that the iterative procedure works only when the energy
is large. This implies that it is a bad approximation at the end of
the generation, when only little energy is remaining. In the \pythia
program this problem is solved by peeling off mesons randomly from
both ends of the string, making two hadrons when the remaining invariant
mass is small enough. This implies that the error from the
``junction'' between the two halfs will be spread out and not visible
in inclusive distributions. This method will, however, not remove the
error on the correlations. They will still remain, as we will discuss
further below.

\subsection{Gluon emission}

The second basic feature of the Lund model is the assumption
that in three dimensions the dynamics of the confined force
field is well represented by the massless relativistic string, 
and that the gluons act as transverse excitations on 
this string \cite{Andersson:1979ij, Andersson:1980vj}. This 
assumption implies angular asymmetries, which
were first observed by the JADE collaboration at the PETRA 
accelerator \cite{Bartel:1981kh}.
A most important feature of this gluon model is
infrared stability. Soft or collinear gluons give 
only small modifications of the string motion, and hence also
on the final hadronic state. 

Although the breakup of a string, which is bent by many gluons, should
also be determined by the area law in eq.~(\ref{eq:breakup}), there
are here ambiguities and technical problems. Is the projection of a
bent string piece onto a meson state the same as for a straight string
piece? The generalization of eq.~(\ref{eq:breakup}) and its
formulation in an iterative process in an event generator also imply
ambiguities \cite{Sjostrand:1984ic}.  Although an error here does not
show up in inclusive distributions, it could well have effects on
correlations, and thus be important in analyses of the FD effects.

\subsection{Baryon production}

Baryon--anti-baryon pairs can be produced when the string breaks by the
production of a diquark--antidiquark pair in a $\bar{3}\!-\!3$ colour
state. The weights must here be adjusted so that the baryons become
fully symmetric spin-flavour states, which also preserves isospin
invariance. This mechanism gives strong correlations between the
baryon and the anti-baryon, which must have two quark flavours in
common. This correlation is stronger than what is observed, and for
this reason we have to imagine that the diquarks can be produced in a
step-like manner, allowing a meson to be produced between the baryon
and the anti-baryon, as indicated in fig.~\ref{fig:popcorn}
\cite{Andersson:1984af}. This so called "popcorn" mechanism also
implies that the baryon and the anti-baryon come farther apart in
rapidity and momentum space.

\FIGURE[t]{
  \includegraphics[angle=0, scale=0.5]{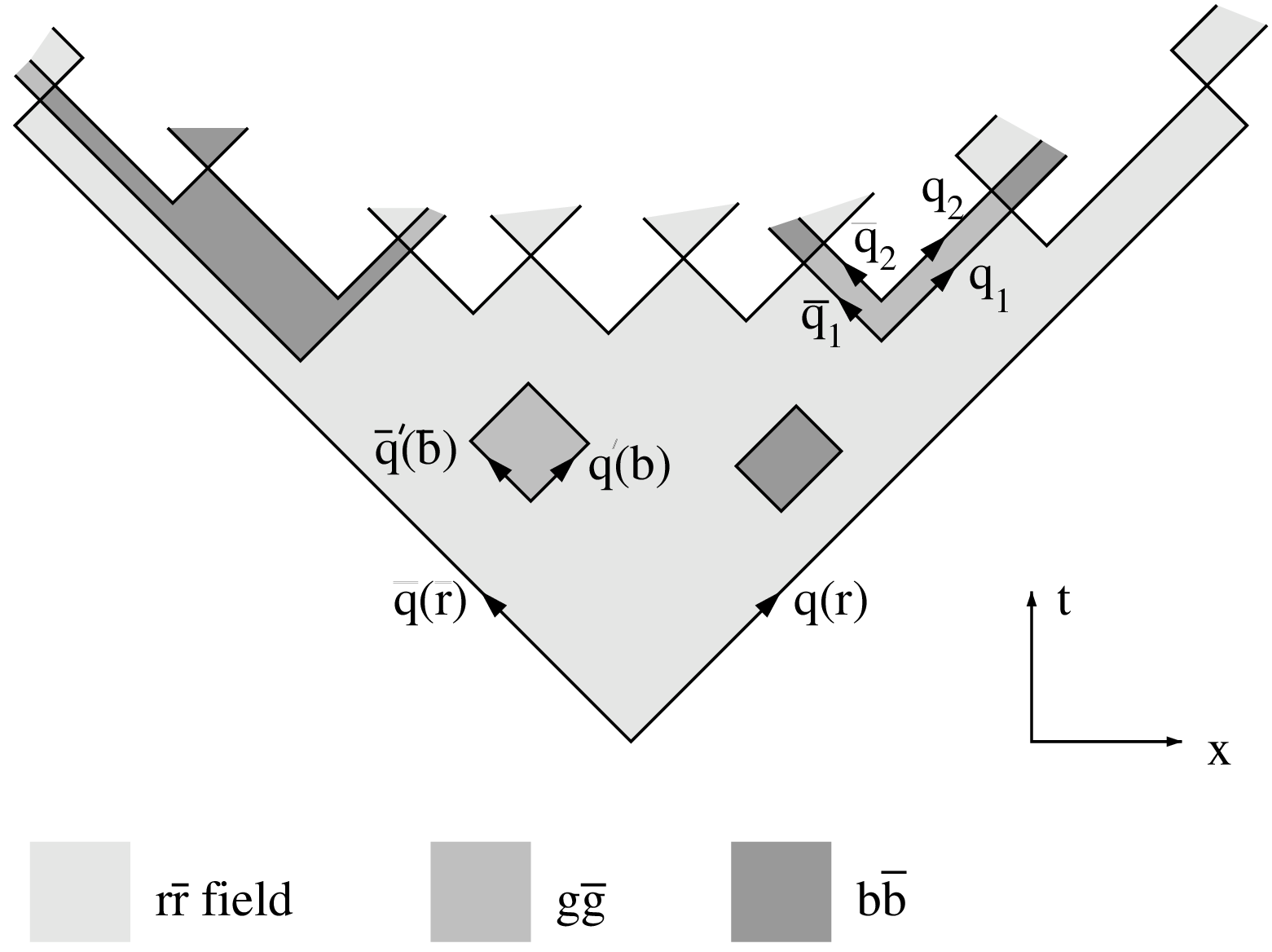}
  \caption{\label{fig:popcorn} Representation of colour fluctuations
    in the string. The wrongly coloured pair
    $q_{1}(b)\bar{q}_{1}(\bar{b})$ together with
    $q_{2}(g)\bar{q}_{2}(\bar{g})$ form an effective
    diquark--antidiquark pair, yielding a baryon and an anti-baryon,
    neighbors in rank.  At the left-hand side we show the production
    of a baryon and an anti-baryon with a meson between them, arising
    from two breakups in a colour fluctuation region.}

}

A more elaborate model for baryon production is developed in 
ref. \cite{Eden:1996xi}. As we find no significant differences
between this model and the standard popcorn model for the 
distributions of interest in this paper, we will not discuss
it further here. 

\subsection{Baryon--anti-baryon correlations}

The ordering of the hadrons along the string, the so called rank 
ordering, agrees on average with the ordering 
in rapidity, with an average separation, $\Delta y$, of the order of half 
a unit in rapidity. As baryon production is suppressed compared to
meson production, a baryon--anti-baryon pair frequently originates from
a single diquark--antidiquark breakup. The baryon and the anti-baryon are
then produced as neighbours in rank, or with one (or a few) mesons in between,
which implies that they are not far away from each other in momentum
space. Two baryons must necessarily come from two different $B\bar{B}$
pairs, and must always be separated in rank by at least one anti-baryon
(and normally also with one or more mesons). This will give a strong
anti-correlation between two baryons in rapidity and in momentum.
This is illustrated in 
fig.~\ref{fig:pp-par}, which shows correlations in $pp$ and $p\bar{p}$
pairs. We see that there is a very strong positive correlation between 
protons and 
anti-protons, which are frequently neighbours in rank, but a negative 
correlation between two protons.

\FIGURE[t]{
  \includegraphics[angle=270, scale=0.4]{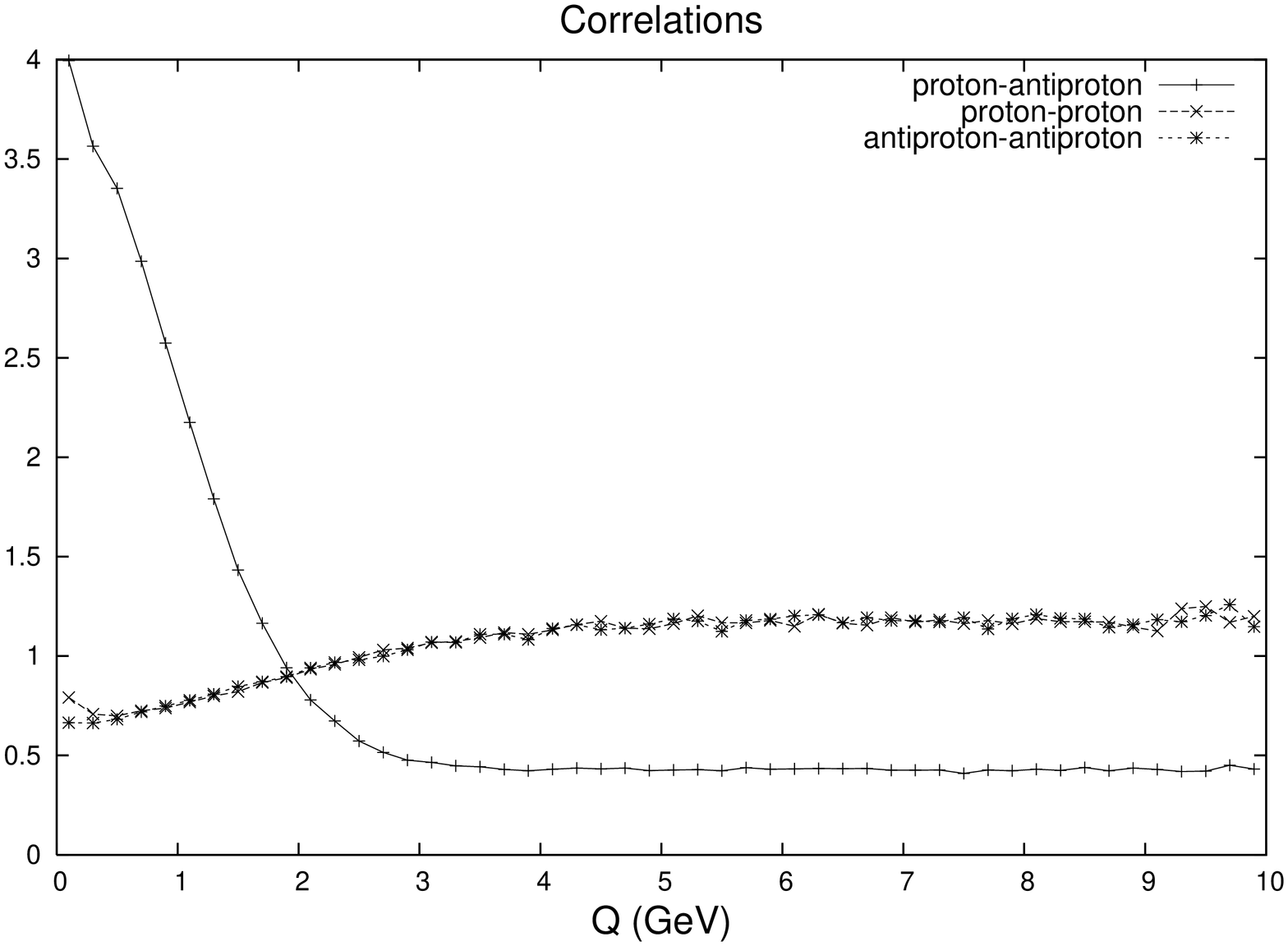}
  \caption{\label{fig:pp-par} Monte Carlo results for the ratio
    $C_{MC}(Q) = MC(Q)/MC_{\mathrm{mix}}(Q)$ for $pp$- ($\times$) and
    $p\bar{p}$-pairs (+).}
}

In fig.~\ref{fig:pp-par} we see that the range for the $pp$ correlation 
is given by $Q \sim 1.5\,\mathrm{GeV} \sim 1/(0.15\,\mathrm{fm})$. We 
note that this corresponds exactly to the correlation length reported in
the experiments. The strength of the correlation is, however, smaller in 
the simulation than in the data.

\emph{This raises the question: Is the difference between data and
  \pythia really a FD effect, or could the event generator
  underestimate the strength of the correlation?}

In the model there is a strong correlation between momentum and space 
coordinates for the produced hadrons. Thus two identical baryons are 
(in the model) well
separated also in coordinate space, and we would from this picture
expect Fermi--Dirac correlations
to correspond to a radius $\sim 2 - 3$~fm. If the dynamical
anti-correlation is indeed underestimated in \pythia, we would in the model
expect that the real FD correlation corresponds to a larger radius, and 
therefore show up for $Q$-values around $1/(2\,\mathrm{fm}) \approx 0.1 
\, \mathrm{GeV}$. Since the phase space is suppressed for these small
$Q$-values, this effect would be impossible to observe in the LEP
experiments. 

Although the fundamental nature of BE correlations is not understood,
its effects have been fairly well reproduced by a model in which the
momenta of the produced hadrons are shifted so that identical mesons
come closer in momentum space \cite{PYBOEI,Lonnblad:1995mr}. This
model is implemented in the program \texttt{PYBOEI} and included in
the \pythia package.  With minor modifications, \texttt{PYBOEI} can
also be used to simulate FD correlations, in which case one would see
an expected drop for very small $Q$-values below 0.5 GeV assuming a
source radius of 1~fm.


\section{Uncertainties in the event generator model}
\label{sec:uncertainties}

There are a number of sources for uncertainty in the \pythia program:
\vspace{1mm}

(1) There are two fundamental parameters, $a$ and $b$, in the
splitting function in eq.~(\ref{eq:f}).
The hadron multiplicity depends essentially on the ratio $(a+1)/b$,
This ratio is therefore well determined by experiments, but 
$a$ and $b$ separately are more uncertain. Small values of $a$ and $b$
correspond to a wide distribution $f(z)$, and a wide distribution in
the separation, $\Delta y$, between hadrons which are neighbours in rank.
Large values of $a$ and $b$ imply a narrow $\Delta y$-distribution
and therefore lower probability for two particles to be close in 
momentum space. The effect of varying $a$ and $b$ keeping the 
multiplicity unchanged is shown in fig.~\ref{fig:abdep}. We see that
larger $a$- and $b$-values give a stronger anti-correlation for
small $Q$.

\FIGURE[t]{
  \includegraphics[angle=270, scale=0.4]{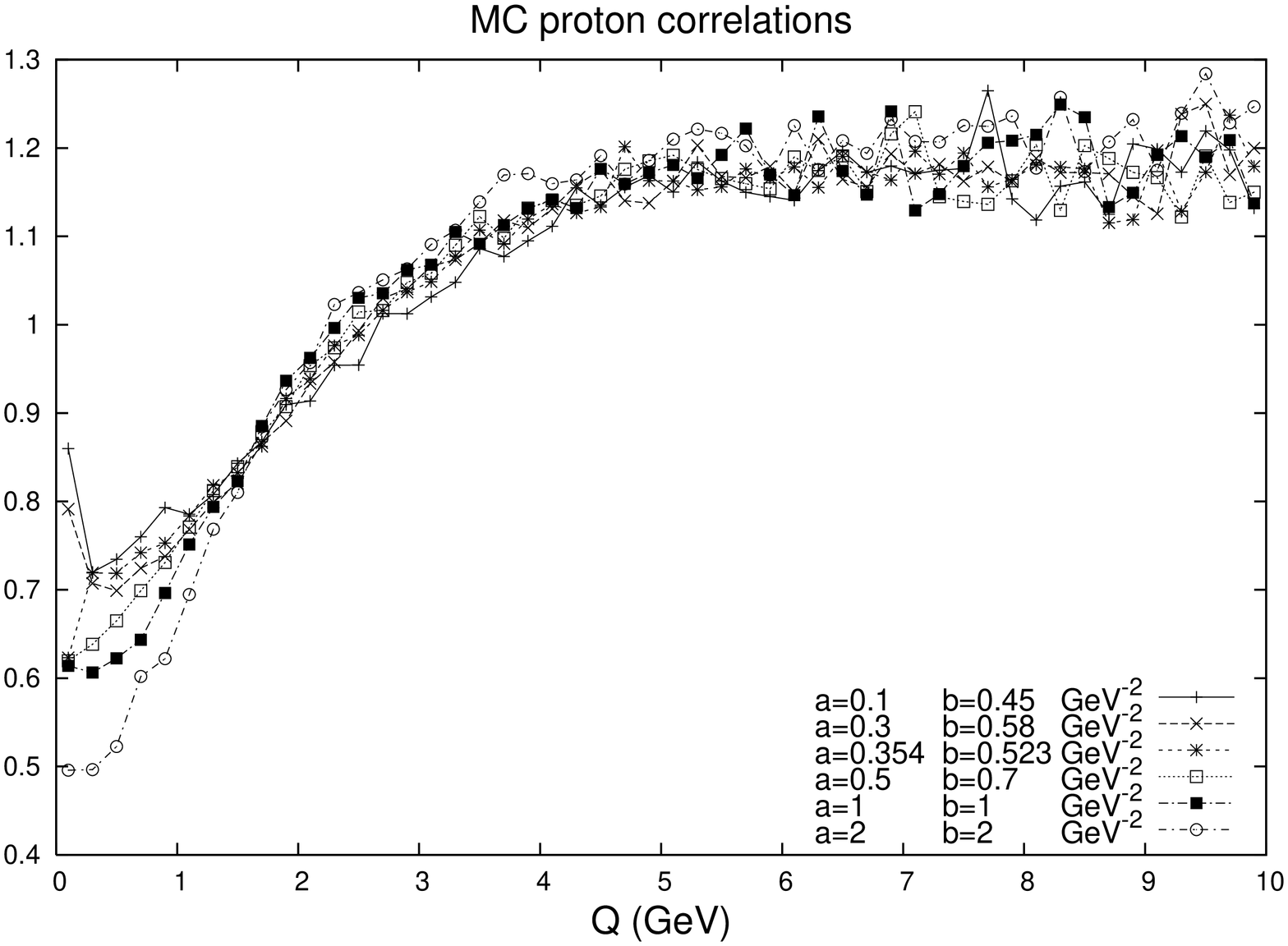}
  \caption{\label{fig:abdep} The ratio $C_{MC}(Q)=
    MC(Q)/MC_{\mathrm{mix}}(Q)$ for different values of the parameters
    $a$ and $b$. Larger $(a,b)$-values give a stronger correlation and
    a deeper dip for small $Q$. The default values in the \protect\pythia
    program is $a=0.3$ and $b=0.58$~GeV$^{-2}$}
}

\FIGURE[t]{
  \includegraphics[angle=270, scale=0.4]{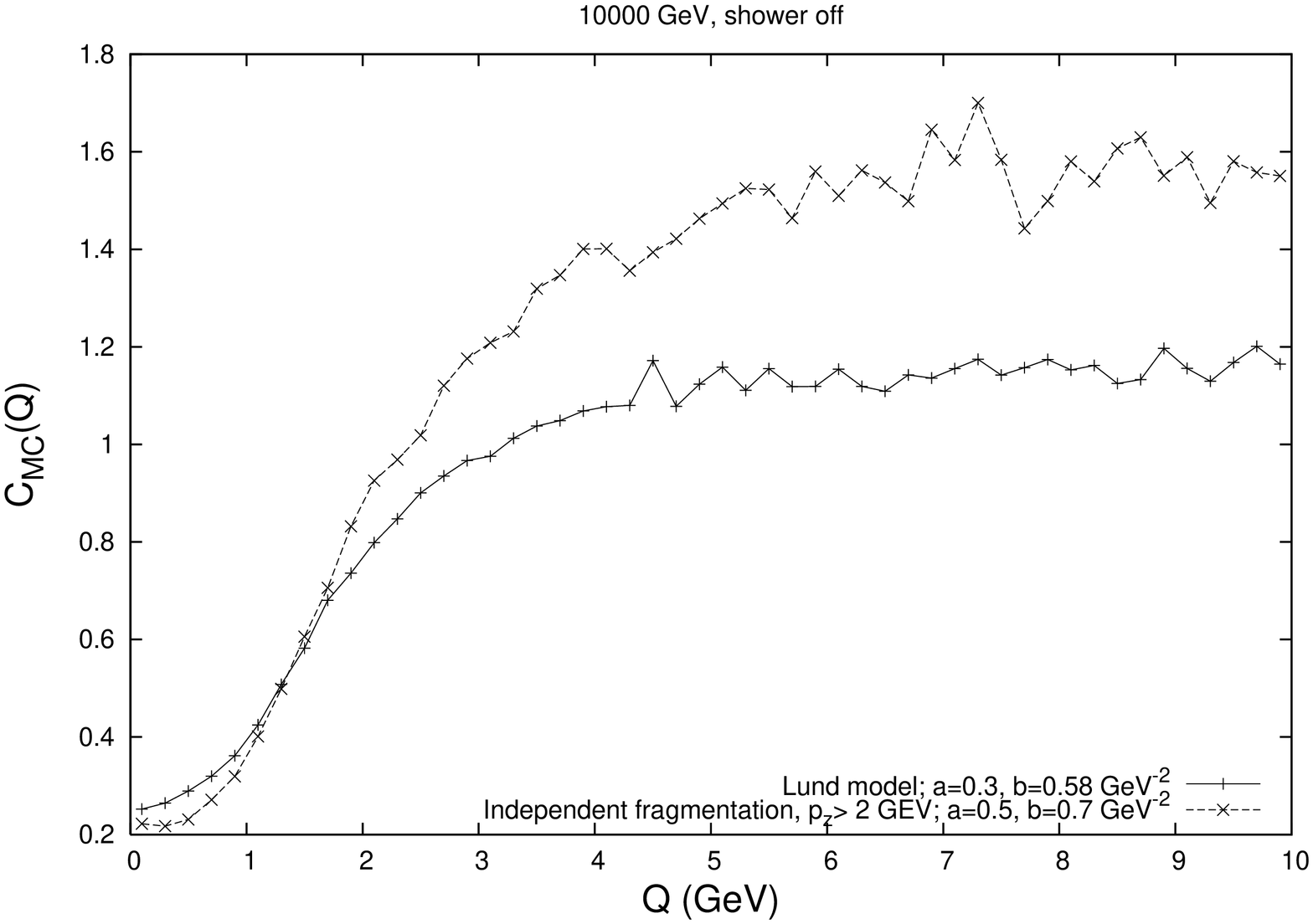}
  \caption{\label{fig:junc} A single jet \emph{without} a junction has
    a large dip in $C_{MC}(Q)= MC(Q)/MC_{\mathrm{mix}}(Q)$ for small
    $Q$ ($\times$).  This correlation is reduced by the approximate
    treatment in the Monte Carlo of the small mass systems close to the
    "junction" (+).}
}

\FIGURE[t]{
  \includegraphics[angle=270, scale=0.4]{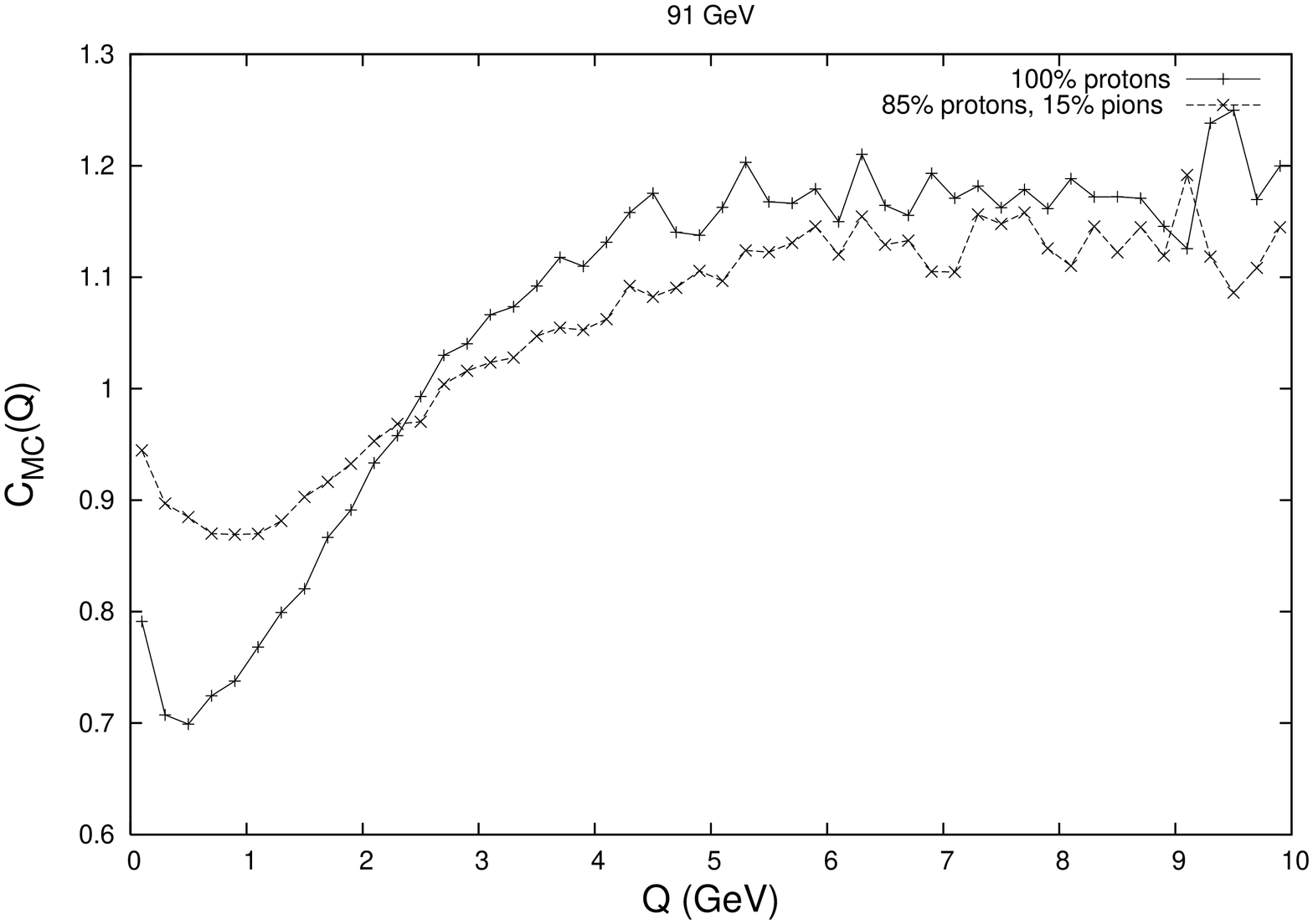}
  \caption{\label{fig:pion} $C_{MC}(Q)= MC(Q)/MC_{\mathrm{mix}}(Q)$
    with (+) and without ($\times$) a 15\% admixture of pions.}
}

(2) The parameter $b$ is a universal constant, but $a$ may be different 
for baryons, although data are well fitted by a universal $a$-value. 
This gives some extra uncertainty.

(3) As discussed in the previous section the \pythia program does not
exactly reproduce the Lund hadronization model.  The splitting
function in eq.~(\ref{eq:f}) gives a correct result when the remaining
energy in the system is large. To minimize the error at the end of the
cascade, when the remaining invariant mass is small, the MC cuts off
hadrons from both ends randomly, and joins the two ends of the system
by producing two single hadrons. It is possible to adjust the cutoff
so that this method works well for inclusive distributions, but it
implies that the correlations do not correspond to the model
prediction for particles close to the ``junction''. To estimate the
error from this approximation we show in fig.~\ref{fig:junc} the $pp$
correlations in a single high energy jet without a junction, and
compare it with the standard result for a 91 GeV $q\bar{q}$-system.
In both cases no gluon radiation is included. We note here that a
significant part of the anti-correlation present in the model has
disappeared in the standard Monte Carlo implementation, as the
approximation in (and around) the junction allows baryons to be
produced with momenta close to each other.

(4) Gluon emissions imply that straight string pieces are small compared to
the mass of a $B\bar{B}B$ system. This gives also extra uncertainty.
The gluon corners on the string imply ambiguities and approximations
in the hadronization model \cite{Sjostrand:1984ic}, which even if not
seen in inclusive distributions may be important for correlations.
This may well be a major reason why $\pi \pi$ correlations are not so well
reproduced by \pythia (see e.g.\ \cite{DeWolf:1992sx, Andersson:1994xd}). 

(5) Besides these model uncertainties, there are uncertainties in the
experimental correction procedure, such as the problems with event
mixing and thrust alignment mentioned in section
\ref{sec:meas-fermi-dirac}.  In addition the identification of
anti-protons is not perfect in the experimental data.  As an example
the D\textsc{elphi} anti-proton sample contains 15\% pions.  As
mentioned above, the pion correlations are not perfectly reproduced by
\pythia.  Fig.~\ref{fig:pion} shows results with and without a 15\%
pion admixture.  We see that an error in the simulation of the
pion--pion or pion--proton correlations also will affect the estimated
proton--proton correlations.

In summary we see that there are many effects which make the Monte
Carlo predictions for $pp$ or $\Lambda\Lambda$ correlations quite
uncertain.  As the experimental determination of the correlations rely
so strongly on a correct Monte Carlo simulation, it is therfore at
present premature to conclude that the production radius has the very
small value around 0.15~fm.  As the expected dynamical $BB$
correlation has the same range ($Q\sim 1.5 \,\mathrm{GeV}$) as the
published results for the FD effect, we believe that it is more likely
that the strength of the dynamical $BB$ correlation is
underestimated in the \pythia event generator. The true FD effect may
then correspond to a larger production region and therefore show up at
$Q$-values too small for experimental observation.

\section{Conclusions}
\label{sec:conclusions}

The reported results on the production radius for baryon pairs is
clearly not consistent with the conventional picture of string
fragmentation. In fact, a production radius of $0.15$~fm, which is
smaller than the size of the baryons themselves seems difficult to
reconcile with any conceivable hadronization model.

There are principal problems in the construction of a reference sample
which contains all dynamical baryon--baryon correlations but not the
effects of Fermi--Dirac statistics. When an event generator is tuned
to inclusive distributions, the use of the double ratio in
eq.~(\ref{eq:BEdouble}) implies that the result depends critically on
a perfect simulation program.

In this article we have noted a number of uncertainties in the Lund
string fragmentation model and its implementation in the \pythia event
generator used in the correction of the data when extracting the
Fermi--Dirac effects. In particular we noted that the strong dynamical
correlation between baryons in the model appears in the same range of
$Q$ as the claimed FD correlations in the data, and that the model
uncertainties for these correlations are large and have not been
independently constrained by data.

In conclusion we feel that it is premature to claim that the observed
discrepancy between data and Monte Carlo is really only due to
Fermi--Dirac effects, which would indicate a new kind of production
mechanism. To make such claims one would first have to demonstrate
that the models correctly describe baryon--baryon correlations in the
absence of FD effects, e.g.\ by comparing model predictions for
$p\Lambda$ and $\bar{p}\bar{\Lambda}$ correlations to experimental
data.

\bibliographystyle{utcaps}  
\bibliography{/home/beckett/leif/personal/lib/tex/bib/references,refs} 
\end{document}